\documentclass[sn-mathphys,Numbered,twocolumns]{sn-jnl}

\usepackage{graphicx}%
\usepackage{multirow}%
\usepackage{amsmath,amssymb,amsfonts}%
\usepackage{amsthm}%
\usepackage{mathrsfs}%
\usepackage{physics}
\usepackage{lmodern}

\usepackage[title]{appendix}%
\usepackage{xcolor}%
\usepackage{textcomp}%
\usepackage{manyfoot}%
\usepackage{multicol}
\usepackage{booktabs}%
\usepackage{algorithm}%
\usepackage{algorithmicx}%
\usepackage{algpseudocode}%
\usepackage{listings}%

\providecommand{\vc}{\vb*}

\begin{document}

\title[]{On-Chip Chiroptical Sensor based on Directional Deflection of Light: A Stern-Gerlach Integrated Optical Analog} 
 
\author[1]{\fnm{Josep} \sur{Martínez-Romeu}}
\author[1]{\fnm{Alejandro} \sur{Martínez}} \email{amartinez@ntc.upv.es}
\author[2]{\fnm{J. Enrique} \sur{Vázquez-Lozano}} \email{enrique.vazquez@unavarra.es}

\affil[1]{\orgdiv{Nanophotonics Technology Center}, \orgname{Universitat Politècnica de València}, \orgaddress{\street{Camino de Vera, s/n Building 8F}, \postcode{46022}, \city{Valencia}, \country{Spain}}}
\affil[2]{\orgdiv{Department of Electrical, Electronic and Communications Engineering}, \orgname{Universidad Pública de Navarra (UPNA)}, \orgaddress{\postcode{31006}, \city{Pamplona}, \country{Spain}}}

\date{\today} 

\maketitle

\begin{abstract}
 CChiroptical techniques for detecting and characterizing the chirality of matter and artificial nanostructures are traditionally based on their interaction with chiral light, typically circularly-polarized fields propagating in free space. More recently, these approaches have been extended to integrated photonic platforms, offering significant practical advantages. However, the generation of chiral guided light is challenging: it requires degeneracy of the fundamental modes of the integrated waveguide, which occurs at a single wavelength and limits broadband characterization. Here, we introduce and simulate numerically a new alternative integrated optical configuration inspired by the Stern–Gerlach experiment, in which a chiral sample is illuminated by a linearly polarized light beam. Such a beam exits from a dielectric waveguide that propagates the fundamental TE mode, thereby eliminating the need for circularly polarized excitation. On this basis, enantio-discrimination is achieved through a spatially resolved scheme using a symmetric arrangement of optical antennas on either side of the propagation axis. Our simulation results suggest that the selective scattering and/or absorption of the circular polarization component matching the chirality of the sample induces an imbalance of (spin) angular momentum of the optical field, resulting in a transverse deflection of the beam toward the corresponding side. This on-chip platform then provides a direct route to chiroptical functionalities under linearly-polarized illumination, enabling compact implementations of chiral sensing, spectroscopy, optical computing, and communication schemes.
\end{abstract}

\section{Introduction}

Chirality is a geometrical property that describes objects that cannot be superimposed with their mirrored images~\cite{Kelvin1904}. It is a fundamental characteristic of nature, appearing in physical, chemical, and biological systems across multiple length scales~\cite{Naaman2019}. From a more pragmatic perspective, chirality plays a particularly crucial role in molecular chemistry, especially within the field of stereochemistry, which studies the spatial arrangement of atoms and its impact on molecular properties~\cite{FDA1992}. In this context, the detection and complete characterization of enantiomers — pairs of molecules with opposite handedness —is of paramount importance, particularly in the chemical and pharmaceutical industries~\cite{Hutt1996,Smith2009}.

Beyond material objects and substances, electromagnetic waves can also exhibit optical chirality~\cite{Tang2010,Tang2011,Bliokh2011}, a property that, like energy or momentum, is a conserved quantity and therefore constitutes a fundamental dynamical property of light~\cite{VazquezLozano2018}. In transverse fields (for example, plane waves and paraxial beams), optical chirality is linked to the degree of circular polarization ~\cite{Forbes2024}. In this sense, circularly polarized light is generally regarded as the paradigmatic example of chiral light~\cite{Schaferling}. In photonics, the detection of chiral substances typically relies on chiroptical interactions mediated by the scattering or absorption of chiral light~\cite{Barron}. For example, the phenomenon known as circular dichroism~\cite{Berova}, refers to the differential absorption of circularly-polarized light with opposite handedness. On this basis, the vast majority of theoretical and experimental proposals for chiroptical devices and applications have been developed~\cite{Schaferling}, both in free-space configurations~\cite{Hendry2010,Tkachenko2014} and, more recently, also suggested in integrated nanophotonic platforms~\cite{VazquezLozano2020,MartinezRomeu2024}.

In this work, we propose and numerically simulate a novel, simplified, and versatile configuration for chiroptical applications in integrated photonics. Specifically, we introduce an on-chip wireless integrated nanophotonic system based on dielectric structures~\cite{GarciaMeca2017,Lechago2018}, aimed at the detection and characterization of chiral particles, as well as at chiral-light routing through scattering-mediated interactions~\cite{Fortuno2015}. The proposed configuration consists of four highly directive waveguides that are adiabatically terminated to act as highly directive in-plane optical antennas, one of them as the illumination path and the other three as output paths, with a common interaction region separating them (see Fig. \ref{Design}). The fundamental TE mode is excited in the input waveguide, so that at its output there is a well-collimated beam with horizontal linear polarization that interacts with a chiral particle placed in the interaction zone, modifying the balance of angular momentum and preferentially directing the scattered light toward one of the two lateral output waveguides. The operating principle relies on the fact that linearly polarized light can be decomposed into two circularly polarized components, in analogy with the decomposition of circular polarization into two orthogonal linear modes. Upon interaction with a chiral particle, one of the circular polarization components is preferentially absorbed, leading to an imbalance in spin angular momentum and, consequently, in the total angular momentum of the light field~\cite{Fortuno2015}. This imbalance results in a transverse deflection of the scattered signal~\cite{Chen2020}, which predominantly couples into one of the lateral output waveguides, thereby emulating the operating principle of an optical analog of the Stern–Gerlach experiment~\cite{Li2007,Kravets2019,Arteaga2019}. As we numerically demonstrate, this angular-momentum–mediated deflection of light brings forth some spectral features showing a direct signature of chirality~\cite{Yoo2015}, forming the basis for the detection, discrimination, and characterization of chiral nanoparticles within an integrated photonic platform under a nonchiral illumination scheme.
\section{Results}

\begin{figure*}[htb!]
    \centering
    \includegraphics[width=\linewidth]{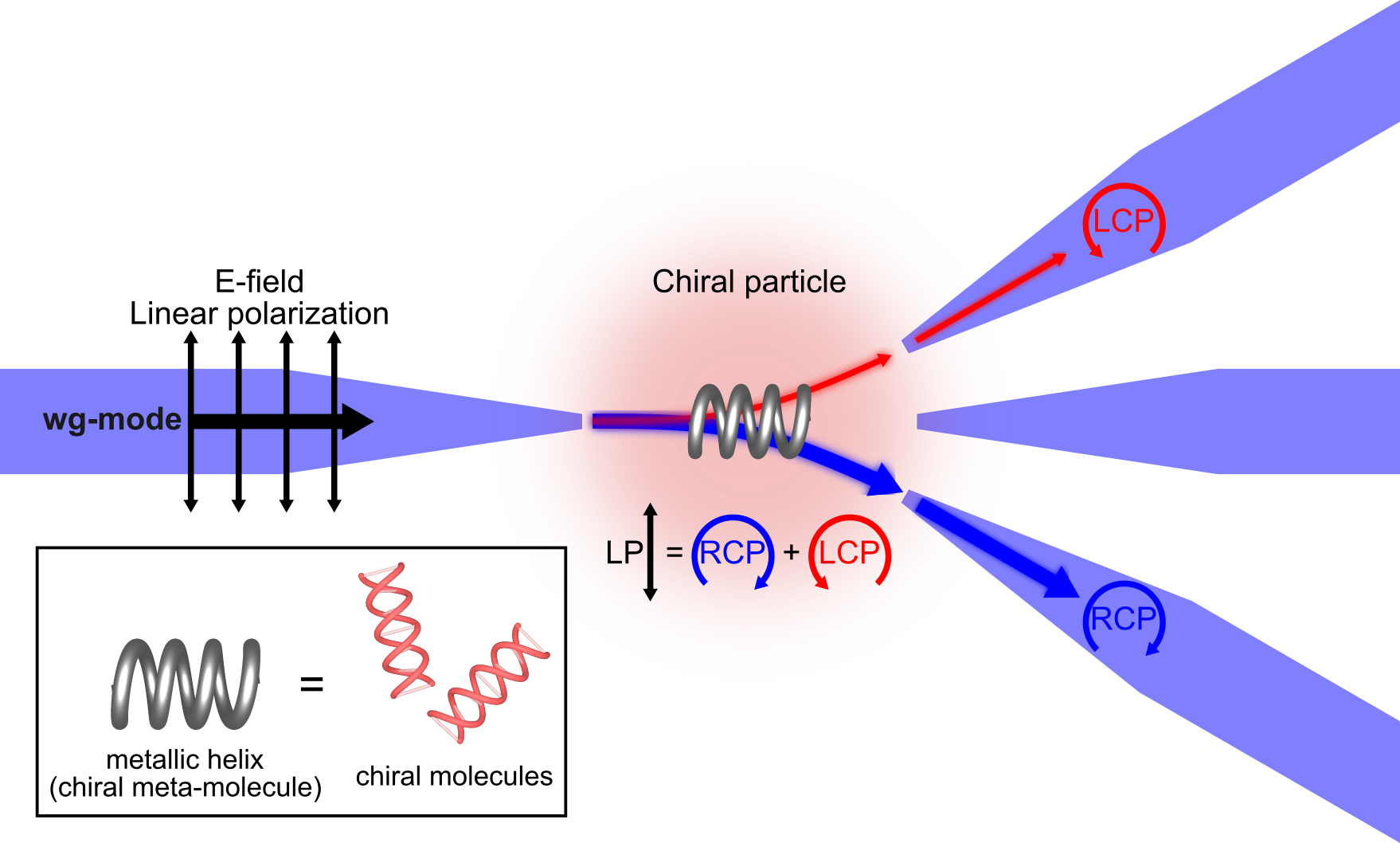}
    \caption{Schematic design of the optical system to perform chiroptical applications based on a configuration similar to the all-optical version of the Stern-Gerlach experiment. In the region between antennas (interaction region), we place a chiral object, which in our simulations is a metallic nanohelix \cite{Hoflich19}. Even though the nature of the chiral response is different, we interpret the nanohelix as a valid analogue to chiral molecules (inset) since it provides stronger - and therefore easier to model and to observe - chiroptical interactions.}
    \label{Design}
\end{figure*}

We consider a system formed by four silicon (Si) waveguides with rectangular cross sections, as shown in the scheme depicted in Fig. \ref{Design}. Other materials could be employed to form the waveguides, as long as they are transparent in the wavelength regime of interest. The integrated waveguides have a rectangular cross-section with 220 nm thickness and 450nm width, which are typical values to achieve single-mode behavior at telecom wavelengths. To maximize the light exiting the input waveguide and entering the output waveguides used to analyze the light scattered by the chiral nanoparticle, the waveguides' terminations include inverted tapers \cite{TaperOL2003} acting as in-plane optical antennas. The tapers begin with a 150 nm width and 220 nm thickness and end with the waveguide cross-section dimensions. The chosen taper length is 4 $\mu$m. This way, we achieve highly-directional transmission from the input waveguide to free space as well as efficient coupling from the scattered field to the output waveguides. Notice that other terminations of the waveguides - or other kinds of optical antennas - could have been chosen to achieve highly collimated beams \cite{GarciaMeca2017,Lechago2018}. 

Following the schematic design shown in Fig.\ref{Design}, we assume that a TE guided mode is injected into the input waveguide, which we term Port 1. This mode has a main component of the electric field in the horizontal plane, and the radiation at the waveguide termination will also be polarized in this direction, as suggested by previous works dealing with on-chip illumination of subwavelength scatterers \cite{AMartinez2016,AMartinez2018,AMartinez2021}. Right in front of the Port 1 taper, we place the chiral component to be considered in our numerical experiment, which, following previous simulations in Ref. \cite{VazquezLozano2020}, we assume to be a metallic nanohelix \cite{Hoflich19}. After the chiral component to be analyzed, we have three more inverted tapers serving as interfaces (antennas) with three output waveguides. The central optical antenna, corresponding to Port 2, is used as a reference but is not actually involved in the detection of chirality. The upward and downward antennas, corresponding respectively to Ports 3 and 4, are set pointing at 30º angles towards the chiral object. These antennas are directly responsible for detecting the object's chirality via differential coupling of the light scattered by the object to each one of them. Indeed, we can directly detect a difference in transmission by measuring the transmission from ports 3 and 4.  We can also measure a dissymmetry factor, which we define as:

\begin{equation}
    A=2\Bigg[\frac{T_4-T_3}{T_4+T_3}\Bigg]
    \label{Dissymmetry2}
\end{equation}

\noindent where $T_3$ and $T_4$ are the transmission efficiencies from Port 1 to Ports 3 and 4, respectively. We use this factor along with the transmittance to test the chiral response of, respectively, a perfect electric conductor (PEC) nanohelix and a silver nanohelix. In both cases, the geometrical chirality causes the structures to interact differently with the incident light depending on its optical chirality. In our simulations, we use the commercial finite-integration technique (FIT) solver in the time domain CST Studio Suite to obtain the spectra of the scattering parameters (S-parameters) for the different output ports. The complete structure, including the waveguides and the nanohelix, was discretized using a hexahedral mesh with 16 cells per wavelength. Additional local refinement was applied in the vicinity of the metallic helix to accurately resolve its subwavelength features. Open boundary conditions were implemented through perfectly matched layers (PMLs) on all external boundaries, with silica assumed as the background medium. The structure was excited via standard waveguide ports at the input. To compute the transmission spectra at the output ports, the supported modes were sequentially excited, and the corresponding S-parameters were recorded.



\begin{figure*}[htb!]
    \centering
    \includegraphics[width=0.7\linewidth]{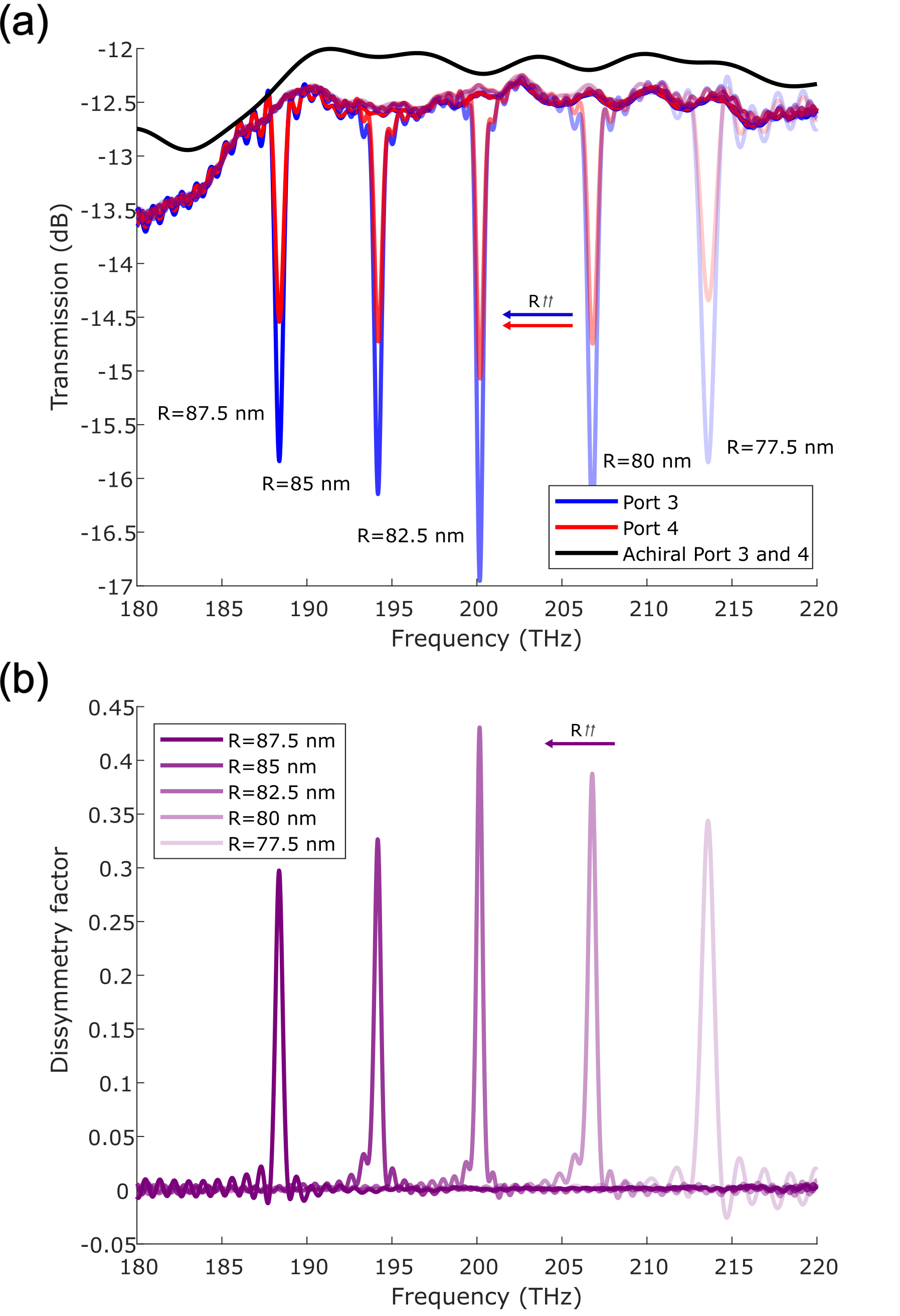}
    \caption{Results with PEC chiral nanohelix.  (a) Transmission spectra for radii 87.5 nm to 77.5nm from left to right. In black, transmission spectra of ports 3 and 4 of an achiral scatterer. (b) Dissymmetry factors for radii ranging from 87.5 nm to 77.5nm. The highest dissymmetry appears for $R$ = 82.5 nm, with a value of 0.43 and at a frequency $f$=200.16 THz. }
    \label{PEC_multi_R}
\end{figure*}

In the first approach, the metal is simply modeled as a PEC. This simplifies the calculations and allows for understanding the basics of the chiral behavior of the helix on account of its geometry itself, dismissing any potential influence of dispersion or absorption effects. In this case, the simulated PEC helix has the following parameters: number of turns $N$=4, total length $L$=240 nm, internal radius $r$=20 nm, and the external radius was varied from $R$=77.5 nm to 87.5 nm in the simulations to check how the variation in the resonance frequency of the nanohelix affects the response, as done previously in Ref. \cite{VazquezLozano2020}. The transmission response towards Ports 3 and 4 is depicted in Fig. \ref{PEC_multi_R}(a), where it can be clearly observed that there are transmission minima, corresponding to the nanohelix resonance, which depend on the geometry of the analyzed particle. Specifically, it is the outer radius of the helix that changes the central frequency of the dips. As would be expected, larger radii have smaller (larger) resonant frequencies (wavelengths). Interestingly, the dips are not symmetric: the transmission dips are between 1 and 2 dB deeper for Port 3 than for Port 4. This is clearly evidenced by the dissymmetry factor shown in Fig. \ref{PEC_multi_R}(b). Due to the symmetries of the whole structure, we attribute the dissymmetry to the chirality of the object, even though it is illuminated by a linearly polarized beam. 

We also analyzed the response of an achiral object for the sake of comparison. The object is composed of four PEC rings having $R$=80 nm, internal radius $r$=20 nm, and distance between the centers of the rings $d$=50 nm made of PEC material. We observe that, for the achiral object, the spectra recorded at Ports 3 and 4 are identical (see black curve in Fig. \ref{PEC_multi_R}(a)) and do not present transmission minima, so the dissymmetry factor is 0 for all frequencies in this case. We also simulated the opposite chirality nanohelix of $R$=80nm and obtained that the dissymmetry factor (not shown) is the exact opposite sign of the one obtained in Fig. \ref{PEC_multi_R}(b).


\begin{figure}[htb!]
    \centering
    \includegraphics[width=\linewidth]{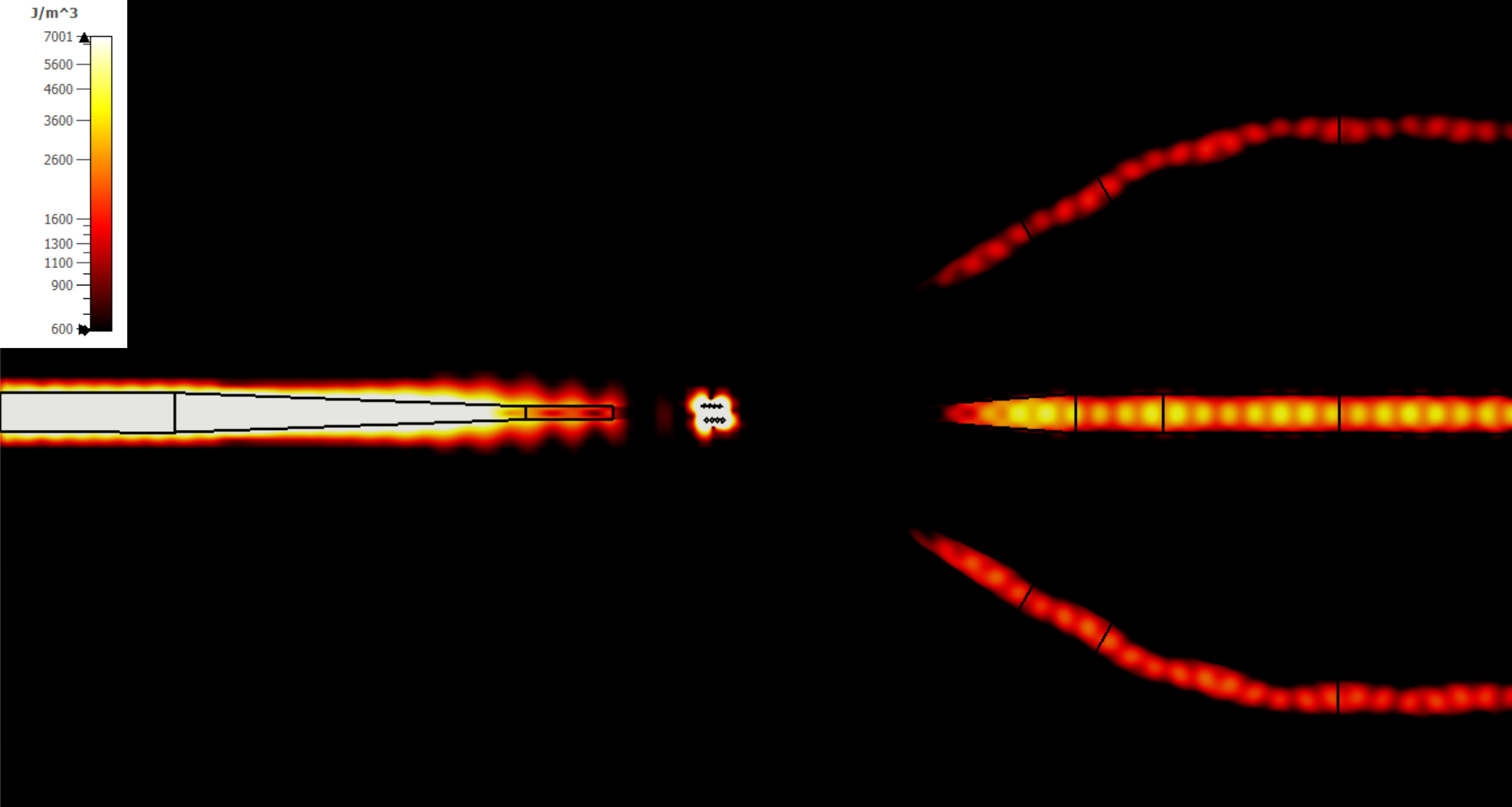}
    \caption{Snapshot of the electromagnetic energy density at frequency 206.8 THz represented in logarithmic color scale. The represented plane cuts through the center of the waveguides and the helix. We can see a higher energy density along Port 4 than along Port 3.}
    \label{We_PEC}
\end{figure}

The analogy between the original Stern–Gerlach experiment and the present configuration can be envisioned as follows: the handedness of the light emerging from the input waveguide (the linear polarization can be considered to be a combination of left- and right-handed circular polarizations) plays the role of the spin of the silver atoms, while the chirality of the helix acts analogously to the magnetic-field gradient that breaks the symmetry and enables spin-dependent beam deflection. We note that related conceptual frameworks have been explored in previous optical realizations of Stern–Gerlach analogs, in which light effectively assumes the role of the deflected particles \cite{Karpa2006}. We simulate a PEC helix of radius $R$ = 80 nm in a resonant frequency of 206.8THz and show this phenomenon in Fig. \ref{We_PEC}. Light emerges from the first antenna and is scattered by the PEC helix. As a direct result of the chiral deflection, while the three nanoantennas receive the incoming light, Port 4 has a larger electromagnetic energy density than Port 3.



To ensure that the nature of the deflection of light is indeed due to the chirality of the object, we study the spin and the helicity of the fields at frequencies for which a nonzero dissymmetry factor is observed. In particular, we calculate the longitudinal component of the spin $\vc{S}_z$ of the field \cite{Golat2023}, which is defined as:
\begin{equation}
    \vc{S}_z=\frac{1}{4\omega}Im(\vc{E}\times\vc{E}^*)_z
\end{equation}
and the optical helicity $\mathfrak{G}$ \cite{Golat2023}, which is defined as:
\begin{equation}
    \mathfrak{G}=\frac{1}{2c_0 \omega}Im(\vc{E}\cdot\vc{H^{*}})
\end{equation}

\begin{figure*}[htb!]
    \centering
    \includegraphics[width=0.7\linewidth]{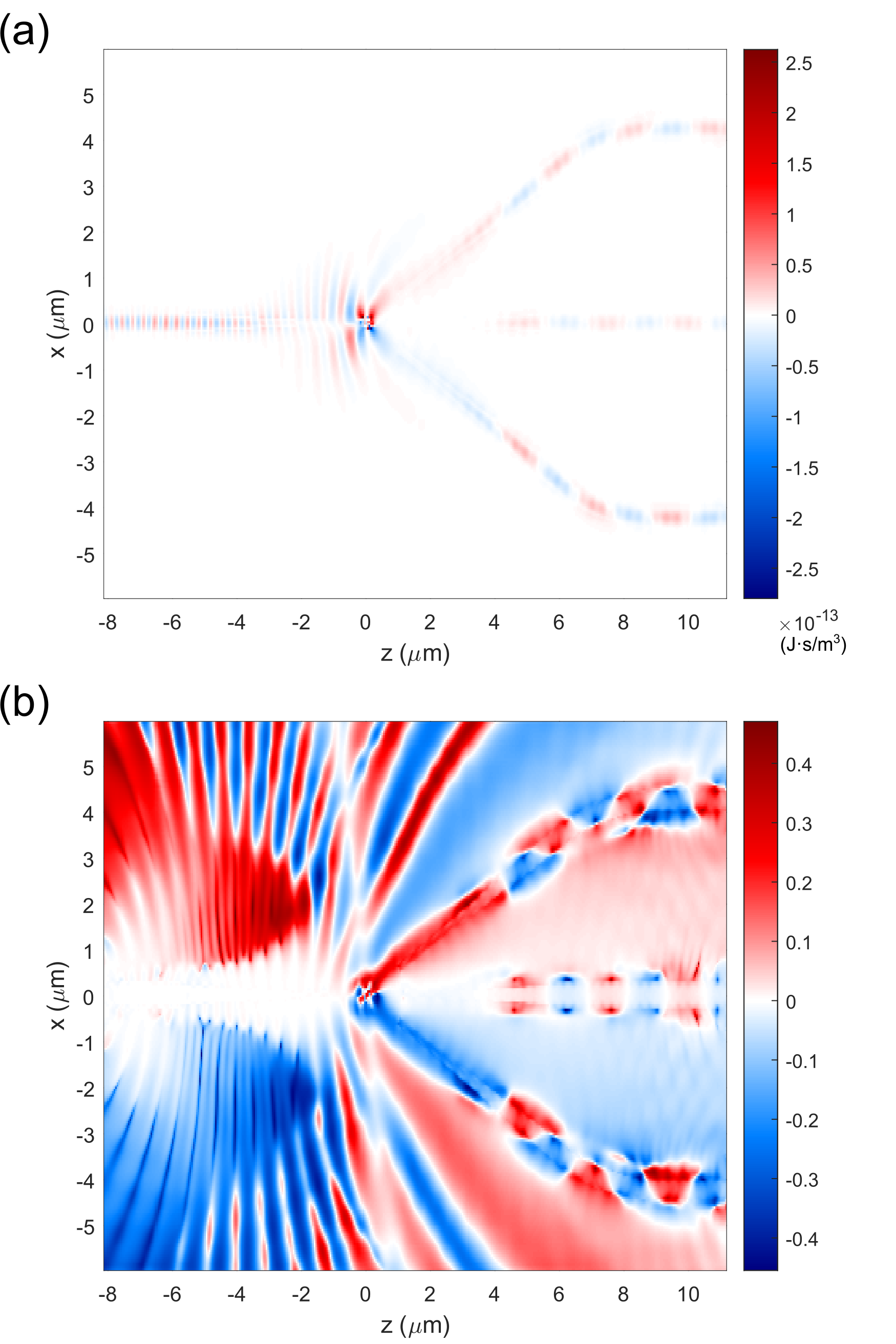}
    \caption{Spin density and normalized helicity density for a PEC chiral nanohelix ($R$=80 nm, $L$=240 nm) at a frequency $f$ = 206.8 THz.  (a)  Longitudinal component of the spin $S_z$. Linearly-polarized light at the output of the exciting antenna illuminates the object, and then chiroptical interaction induces an asymmetric, non-zero $S_z$. After interacting with the chiral scatterer (in the center) the deflected light has positive spin when coupled to the optical antenna corresponding to port 3 and negative spin light couples to port 4. (b) Normalized helicity defined as $\mathfrak{G}/\sqrt{W_e W_m}$. Where $W_e=\frac{\varepsilon}{4}|\vc{E}|^{2}$ and $W_m=\frac{\mu}{4}|\vc{H}|^{2}$. We observe an equivalent effect as in the spin density: Light coupling to the waveguide going to port 3 has positive helicity, while the light coupling to port 4 has negative helicity. }
    \label{fig:spin_helicity}
\end{figure*}

Figures \ref{fig:spin_helicity}(a) and (b) show, respectively, the longitudinal spin and the optical helicity (normalized to the energy density for clarity reasons in the plot) for the chiral object on the plane $y$=0. From these results, we interpret that the light is deflected analogously to the Stern-Gerlach case, showing opposite helicities along each path, which we attribute to chiroptical interaction in the chiral object.  Indeed, we repeated our calculations but for the response of the achiral object described above and obtained that both the spin and helicity were zero over the entire simulation domain.

\begin{figure*}[htb!]
    \centering
    \includegraphics[width=0.7\linewidth]{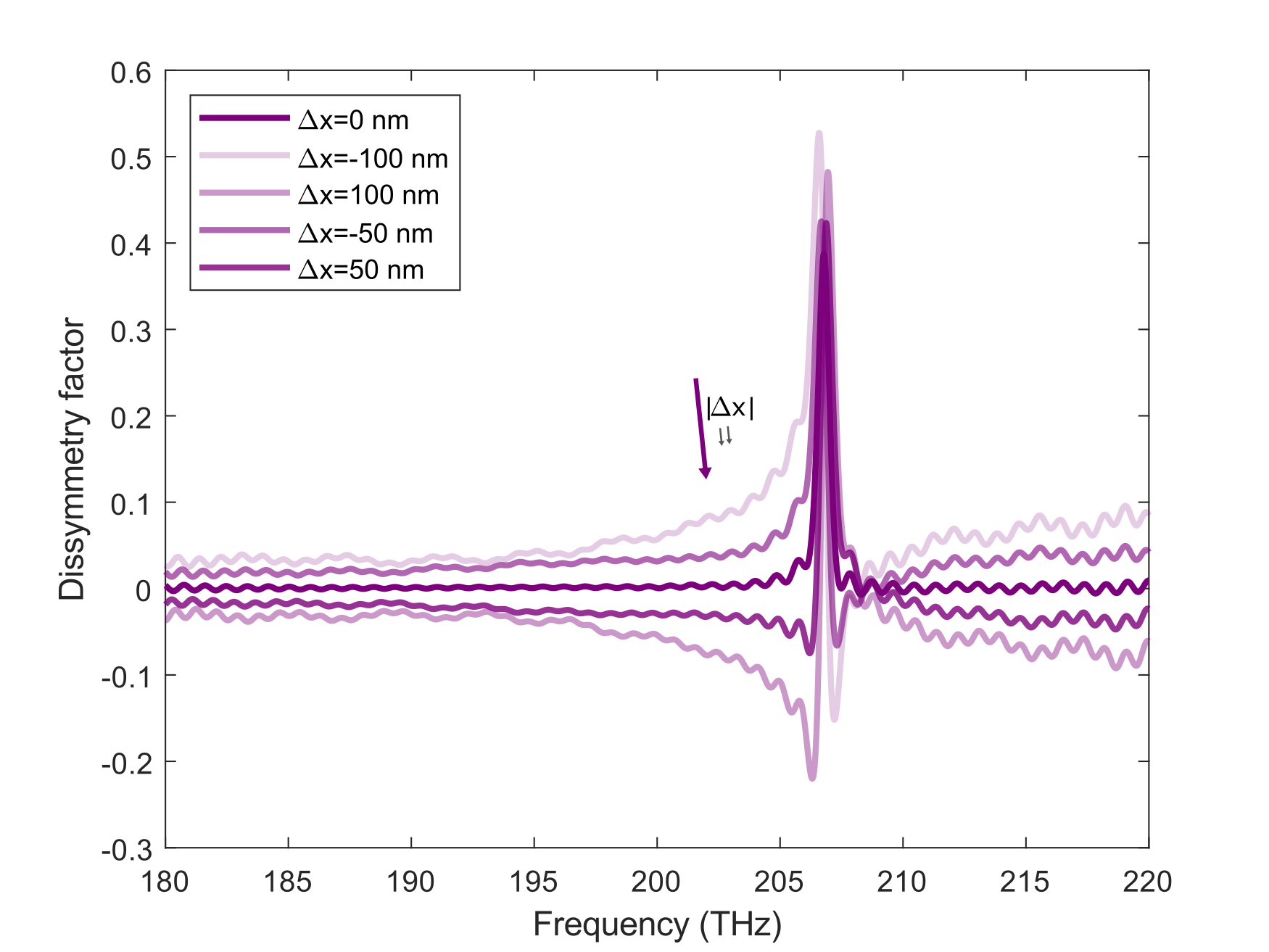}
    \caption{Dissymmetry factor for a PEC chiral nanohelix ($R$=80 nm, $r$=20nm, $L$=240 nm) under lateral displacements along the $x$-axis. $\Delta x$ corresponds to the displacement from the initial position of the helix with respect to the x axis as defined in Fig. \ref{fig:spin_helicity}}
    \label{fig:displacements}
\end{figure*}

We also test the robustness of the measurement setup by displacing the nanohelix in the $x$ axis and calculating its response. We observe that, besides the peak in the dissymmetry factor appearing for all the spatial displacements, and whose sign is kept in all the cases, there is also a non-negligible difference in transmission appearing outside the peak when $\Delta x$ $\neq $ 0. This tail in the dissymmetry can be attributed to the asymmetry of the structure, but the overall effect remains even for short displacements.

\begin{figure*}[htb!]
    \centering
    \includegraphics[width=0.7\linewidth]{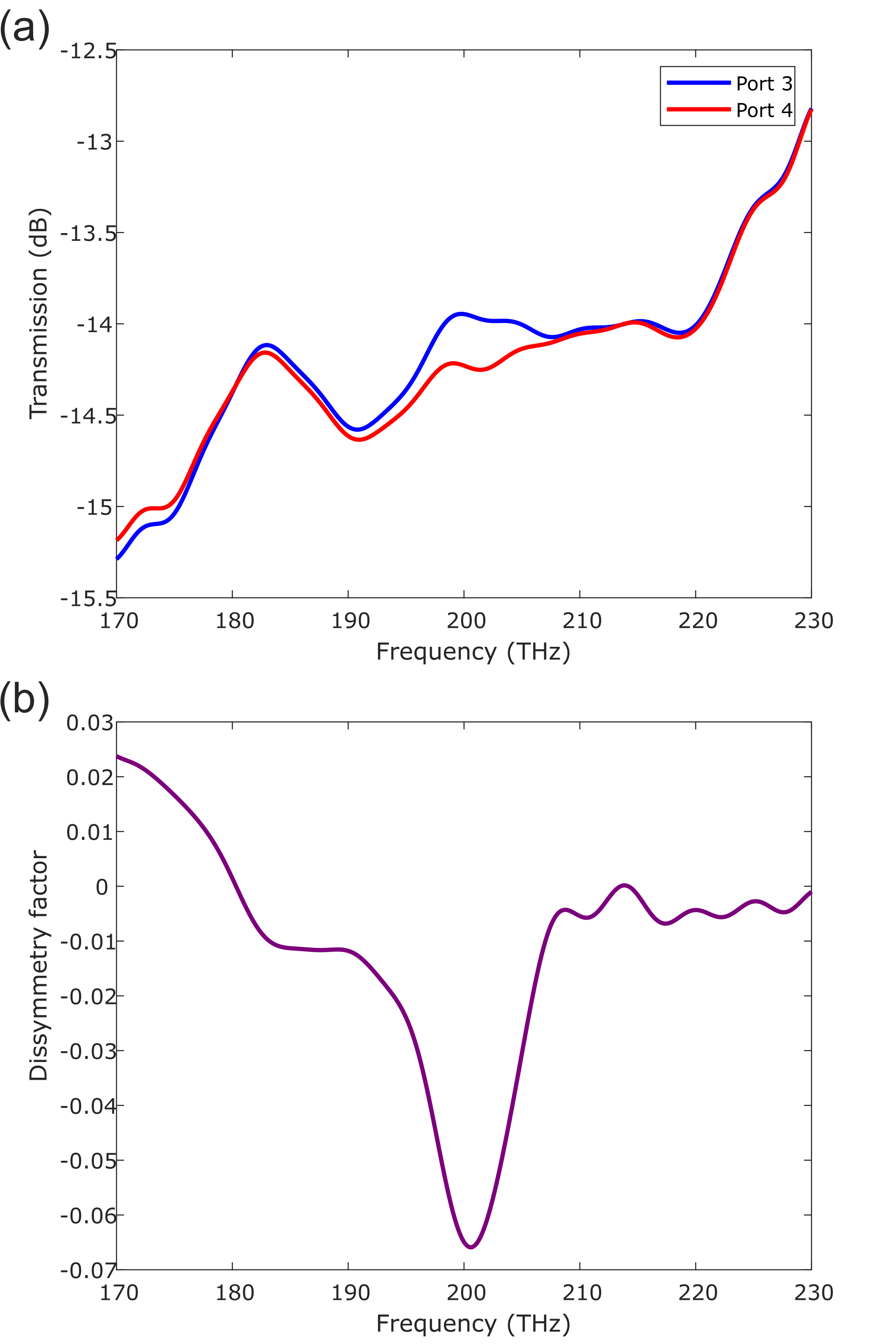}
    \caption{Results for a silver chiral nanohelix ($R$=80 nm, $L$=280 nm).  (a) Transmission spectra corresponding to Ports 3 and 4. (b) Dissymmetry factor calculated with Eq. \ref{Dissymmetry2}. }
    \label{Silver}
\end{figure*}

In the next step, we performed simulations considering that the nanohelix placed in the interaction region is made of silver to account for absorption of the material in the overall response. To this end, we model the silver using parameters from Ref.~\cite{Johnson1978}. In this case, we consider that the silver chiral nanohelix has a number of turns $N$ =4, a total length $L$=280 nm, an internal radius $r$=10 nm, and a single value for the external radius $R$=160 nm. The obtained results are represented in Fig.~\ref{Silver}. The transmission response, shown in Fig.~\ref{Silver}(a), shows that the most pronounced difference in transmission between Ports 3 and 4 takes place at the frequency bandwidth range from 191 THz to 208 THz. In Fig.\ref{Silver}(b), we observe the maximum magnitude of the dissymmetry factor to be $A$=-0.0658 found at 200.54 THz. Remarkably, the inclusion of material absorption broadens the operational bandwidth and reduces the contrast between the output ports. This could be expected, as absorption losses tend to broaden and make less pronounced the response of an optical resonator. Still, the chirality-induced asymmetry remains clearly observable and, compared to the PEC case, persists over a wider frequency~range, so our results are consistent with expected resonance behaviour for realistic materials.

\begin{figure*}[htb!]
    \centering
    \includegraphics[width=0.7\linewidth]{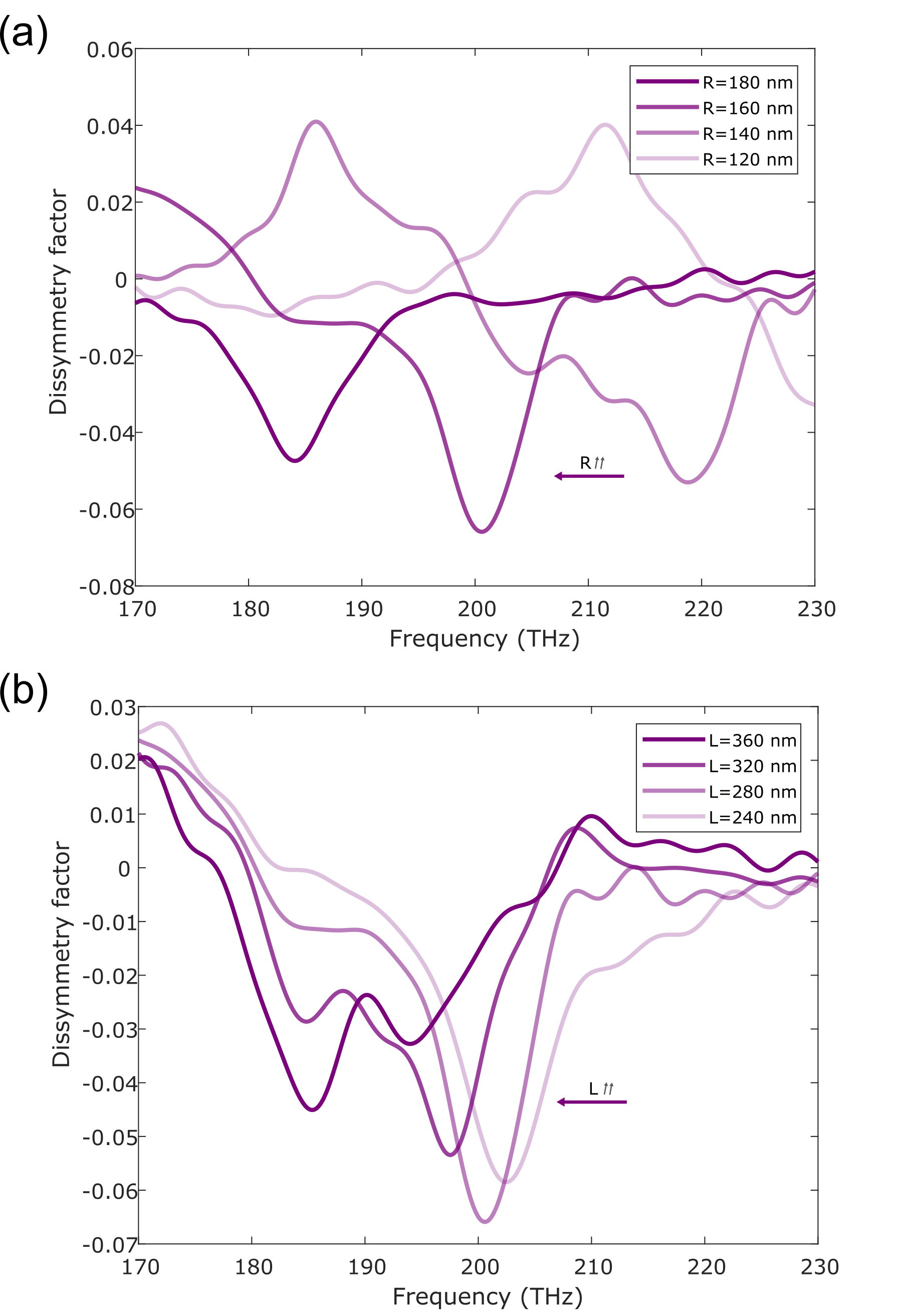}
    \caption{ Dissymmetry factors of the silver nanohelix for different radii (a) and lengths (b). }
    \label{Silver_L}
\end{figure*}

Finally, we study the response of the silver structure when changing two parameters of the helical analyte. We have changed the external radius while maintaining the total length as $L$=240 nm in Fig.~\ref{Silver_L}(a), and we have changed the total length while maintaining the external radius $R$ = 160 nm in Fig.~\ref{Silver_L}(b). In Fig.~\ref{Silver_L}(a), we can see that larger radii present smaller resonant frequencies as expected from the results obtained with PEC in Fig.~\ref{PEC_multi_R}. Another remarkable observation is that the shape of the spectrum is more significantly altered than it was for PEC. This suggests that for realistic structures, the mechanisms for chiral dissymmetry present a combination of material absorption and size parameters. On the other side, in Fig. \ref{Silver_L}(b), we can see that nanohelices with different lengths have different central peak frequencies. Longer nanohelices have smaller central peak frequencies. The central frequency shift is non-linear with respect to the length of the nanohelix. This contrasts with the behavior observed with respect to the radius of the nanohelix and might have interest for future characterization of chirality in modeled objects.

\newpage

\section{Conclusions}

In this work, we have proposed and demonstrated using numerical simulations a novel design inspired by an optical analog of the Stern–Gerlach configuration for chiroptical applications based on integrated photonic platforms. In the proposed scheme, the chirality under investigation corresponds to that of a sample illuminated by a linearly polarized light beam, enabling spatial enantio-discrimination without the need for circularly polarized excitation. Specifically, a set of optical antennas, implemented as inverted tapers and placed symmetrically on the left and right sides of the propagation axis, allows the spatial separation of chiral responses. The selective scattering of the circular polarization component matching the sample chirality generates an imbalance in angular momentum, which results in a transverse deflection of the beam toward the corresponding side. 

Building on this physical mechanism, we have numerically demonstrated the feasibility of the proposed configuration for chiral sensing applications, enabling the detection and characterization of chiral nanoparticles. As a first proof of concept, we considered PEC chiral nanoparticles, for which the interaction is governed solely by the geometric features of the chiral structures. We then extended the analysis to silver nanoparticles to assess the impact of realistic material dispersion and absorption effects. Future work will focus on optimizing the device geometry and antenna arrangement, as well as on simplifying and enhancing the measurement schemes. 

Overall, the proposed integrated photonic platform constitutes a lab-on-a-chip–based approach capable of inducing chiroptical interactions between chiral substances under nonchiral illumination. This paradigm opens new avenues for the development of integrated applications in chiral sensing, spectroscopy, enantioselectivity, optical computing, and optical communications.

Data underlying the results presented in this paper are available in \cite{Dataset}.

\end{document}